# Decoherence and coherence in gravitational, electric and strong nuclear fields

P. R. Silva – Departamento de Física  - ICEx – Universidade Federal de Minas Gerais

C. P. 702 – 30123-970 – Belo Horizonte – MG - Brazil

e-mail: prsilvafis@terra.com.br

**ABSTRACT**- Inspired in the work of Erich Joos which appreciated the role played by matter in making the decoherence of the gravitational field, we developed an alternative way of treating the former problem. Besides this, we used the alternative approach to examine the decoherence of the electric field performed by the conduction electrons in metals. As a counterpoint, we studied the coherence of the electric color field inside nucleons, which renders the strong field a totally quantum character.

## 1 –    Introduction

In the paper: "Why do we observe a classical spacetime?", Joos [1] (please see also [2] and references cited in both papers) analyses the interaction of the gravitational field with the matter represented by a cubic centimeter of air at STP conditions and concludes that: Thus the matter content of the universe not only "tells space how to curve", but also "tells space to behave classically".

In his work [1,2] Joos proposes that the quantum behavior of the gravitational field does not happen only near the Planck scale. As pointed out by Joos [1,2], the gravitational field, i. e. spacetime, is continuously measured by matter. This measurement destroys interferences between different field strengths and thereby renders gravitation a classical entity. In quantum gravity there is not classical spacetime at starting, when the theory is considered at the most fundamental level. Since it is supposed that the gravitational field has to be quantized by consistency reasons, an important question to be answered is how the emergency of the classical properties is achieved.

In this note we want to discuss the Joos [1,2] proposal in a slightly  different way. Besides to examine the decoherence of the gravitational field by the matter, as in Joos paper [1], we also address to the problem of the decoherence of the electric field by the electrons of conduction in a metal.  Finally we use the present formalism to investigate the coherent regime of the strong nuclear field.

## 2 –    Joos' derivation

Let us make a sketch of the Joos [1] derivation. He considered a homogeneous gravitational field which feels a cube of length L in a quantum state $\Psi$ at some initial time $t = 0$

$$|\Psi> = c_1|g_1> + c_2|g_2>, \qquad (1)$$

being $g_1$ and $g_2$ two different field strengths. A particle of mass m in a state $|\chi>$ which travels through this volume is sensible to the different values of g, once its trajectory is influenced by them:

$$|\Psi\rangle |\chi^{(0)}\rangle \rightarrow c_1|g_1\rangle |\chi_{g1}(t)\rangle + c_2|g_2\rangle |\chi_{g2}(t)\rangle. \tag{2}$$

This correlation destroys the coherence between $g_1$ and $g_2$, and the reduced density matrix after many of such interactions, is supposed to take the form

$$\rho(g_1, g_2, t) = \rho(g_1, g_2, 0) \exp[-\Gamma t \,(g_1 - g_2)^2], \tag{3}$$

where

$$\Gamma = n\, L^4 [(\pi m)/(2k_B T)]^{3/2}, \tag{4}$$

for an ordinary gas with a particle density n and temperature T. For example, air under ordinary conditions (STP), L=1cm, and t =1s yields a remaining coherence width of [1,2]

$$\Delta g / g \approx 10^{-6}. \tag{5}$$

## 3 –   An alternative way to Joos' derivation

Giving the two possible values to be assumed by the quantum gravitational field, namely $g_1$ and $g_2$, we are led to think the problem as a two-level system, where the interaction between the field and the gas particle is described by a Dirac-like field satisfying the first-order differential equation

$$(c^{-1})\, \partial\Psi/\partial t + \partial\Psi/\partial x = [(m\, \Delta g\, L)/(\hbar c)]\, \Psi - [(mv)/\hbar]\, |\Psi^*\Psi|\, \Psi. \tag{6}$$

In (6) $\Delta g = |g_2 - g_1|$, v is the characteristic speed of a gas molecule, $\hbar$ is the reduced Planck constant and c is the speed of light in vacuum.

A solution of (6) is

$$\Psi = \Psi_1 \exp[i\,(k\,x - \omega t)], \tag{7}$$

where

$$\Psi_1^{\,2} = (\Delta g\, L)/(vc). \tag{8}$$

Meanwhile by using the Heisenberg uncertainty principle we can write

$$(m\, \Delta g\, L)\, \Delta t = h, \tag{9}$$

which leads to

$$1/(\Delta t) = (m\, \Delta g\, L)/h = \upsilon. \tag{10}$$

Then, the line width due to a one-particle scattering is given by

$$\gamma_1 = \upsilon\, \Psi_1^{\,2} = [m\,(\Delta g)^2\, L^2]/(hvc). \tag{11}$$

Now we define for the one-particle density matrix $\rho_1$, the potential $F_1$, namely

$$F_1 = \tfrac{1}{2}\, \gamma_1\, \rho_1^{\,2} + \ldots, \tag{12}$$

where $\rho_1$ corresponds to $\Psi_1^{\,2}$ given by (8). We propose that $\rho_1$ satisfies a kind of Landau-Khalatnikov equation

$$\partial \rho_1 / \partial t = -\, \partial F_1 / \partial \rho_1. \tag{13}$$

By considering only the first term of the expansion (12), the solution of (13) yields

$$\rho_1(t) = \rho_{01} \exp(-\gamma_1 t). \tag{14}$$

The N-particle density matrix can be written as

$$\rho = \prod \rho_1 = (\rho_{01})^N \exp(-N \gamma_1 t). \tag{15}$$

We obtain from (15)

$$\gamma = N \gamma_1 = \tau^{-1}. \tag{16}$$

Inserting (11) into (16) and solving for $\Delta g$, we get

$$\Delta g = [(hvc)^{1/2} \tau^{-1/2}] / [L (Nm)^{1/2}]. \tag{17}$$

Putting $\tau = 1s$, $L = 1$ cm, $v = 3.5 \times 10^2$ m/s, and $m = 4 \times 10^{-26}$ Kg into (17), yields

$$\Delta g / g |_{\text{this work}} = 10^{-7}, \tag{18}$$

which is an order of magnitude smaller than the result obtained by Joos [1,2] (please see (5)).

In order to better compare this work with Joos' results it is convenient to define $\Gamma_{\text{new}}$ as

$$\gamma = \Gamma_{\text{new}} (\Delta g)^2 = [N L^2 m (\Delta g)^2] / (hvc). \tag{19}$$

Therefore

$$\Gamma_{\text{new}} = (NmL^2) / (hvc) = (nL^5 m) / (hvc). \tag{20}$$

Taking

$$v = [(8 k_B T) / (\pi m)]^{1/2}, \tag{21}$$

we get

$$\Gamma_{\text{new}} = (n L^5 m^{3/2} \pi^{1/2}) / [(hc(8 k_B T)^{1/2}]. \tag{22}$$

The above result must be compared with that obtained by Joos, given by eq. (4) of the present work. We see that $\Gamma_{\text{new}}$ of equation (22) shows an explicit dependence on the Planck constant h, and c, the light speed in vacuum.

## 4 – Decoherence of the electrical field inside metals

The electrical conductivity in metals can be considered as the response of the electrons of the conduction band to an applied electric field. Let us assume as in the previous case this field as being the superposition of two quantum-field states, namely $E_1$ and $E_2$. Again, thinking in terms of a two-level system, we can write a first order differential equation describing the interaction of the electric field with the matter, represented by the free electrons traveling at the Fermi velocity.

We write

$$(c^{-1}) \partial\Phi / \partial t + \partial\Phi / \partial x = [(e \Delta E L) / (\hbar c)]\Phi - [(m v_F) / \hbar]|\Phi^*\Phi|\Phi. \tag{23}$$

In (23), $\Phi$ is a Dirac-like field, $\Delta E = | E_2 - E_1 |$, L is the lateral length of a cube, m the electron mass and $v_F$ its Fermi velocity, and e is the quantum of electric charge.

A solution of (23) is

$$\Phi = \Phi_1 \exp[i(kx - \omega t)], \tag{24}$$

where

$$\Phi_1{}^2 = (e \, \Delta E \, L) / (m \, v_F \, c). \tag{25}$$

From the uncertainty principle, we also obtain

$$\upsilon = 1/(\Delta t) = (e \, \Delta E \, L)/h. \tag{26}$$

The one-particle line width is

$$\gamma_1 = \upsilon \, \Phi_1{}^2 = [e^2 \, (\Delta E)^2 \, L^2]/(h m \, v_F \, c). \tag{27}$$

Following steps which go from (12) to (16), we obtain for the N-particle line width

$$\gamma = N \, \gamma_1 = \tau^{-1} = N\{[e^2 \, (\Delta E)^2 \, L^2]/(h m \, v_F \, c)\}. \tag{28}$$

Finally solving for $\Delta E$, we get

$$\Delta E = [(h m v_F \, c)^{1/2}] / [e \, L (N \, \tau)^{1/2}]. \tag{29}$$

To give a numerical example: The electric field inside a copper wire with a current density of $5.1 \times 10^5$ A/m$^2$, is of order of magnitude of $10^{-2}$ V/m [3]. Taking L = 1cm, N = $9 \times 10^{22}$, the mean free time (at the room temperature) $\tau \approx 10^{-14}$s, and $v_F = 1.6 \times 10^6$ m/s, we have

$$\Delta E = 10^{-9} \text{ V/m}. \tag{30}$$

Therefore (after this extremely short time interval) the value of E is well defined up to

$$\Delta E / E \approx 10^{-7}. \tag{31}$$

## 5 –    Color field inside the nucleon

If we assume that strong interaction gives nucleon its mass, we may use the formalism applied before to the gravitational and electric field cases as a means to infer about some features of the electric color field $\mathcal{E}$. Let us write

$$(c^{-1}) \, \partial\Theta/\partial t + \partial\Theta/\partial x = [(q \, \Delta\mathcal{E} \, L)/(\hbar c)] \, \Theta + [p/\hbar] \, |\Theta^*\Theta| \, \Theta. \tag{32}$$

In (32) $\Theta$ is a Dirac-like field, q is the color charge, $\Delta\mathcal{E} = |\mathcal{E}_2 - \mathcal{E}_1|$ is the difference between the two eigen-values of the color field $\mathcal{E}$, and p is a characteristic relativistic momentum to be evaluated in the following, and L is some length scale of the nucleon.

In the MIT bag model of the nucleon [4,5,6] the pressure B of vacuum over the boundary of the bag volume V is related to the nucleon mass-energy by

$$B \, V = (1/4) \, m_n \, c^2 = (1/4) \, \rho V \, c^2. \tag{33}$$

In (33), $\rho$ is the averaged mass density of the nucleon.

The velocity v of the "sound" propagating in the nucleon matter is then given by

$$v = (B/\rho)^{1/2} = (1/2) \, c, \tag{34}$$

and the relativistic momentum of a particle of mass $m_n$ and velocity $c/2$ is

$$p = (m_n \, c)/(3)^{1/2}. \tag{35}$$

A solution of (32) can be obtained in an analogous way of the solutions of (6) and (23), with the amplitude squared

$$\Theta_1{}^2 = [(q \, \Delta\mathcal{E} \, L \, 3^{1/2})/( \, m_n \, c^2)]. \tag{36}$$

From the uncertainty principle we get

$$1/(\Delta t) = \upsilon = (1/h)(q\, \Delta\mathcal{E}\, L). \tag{37}$$

Then by considering the one-particle scattering we have

$$\tau^{-1} = \upsilon\, \Theta_1^2 = [(q\, \Delta\mathcal{E}\, L)^2\, 3^{1/2}]/(h\, m_n\, c^2). \tag{38}$$

Solving for $\Delta\mathcal{E}$, yields

$$q\, \Delta\mathcal{E}\, L = (\tau)^{-1/2}\, c\, (h m_n)^{1/2}\, 3^{-1/4}. \tag{39}$$

Now, for the averaged value of the color field it is possible to write

$$q <\mathcal{E}> L = m_n\, c^2. \tag{40}$$

It seems that the "maximum fluctuation" of the color electric field will occur, when its width $\Delta\mathcal{E}$ equals to its averaged value. Therefore making the equality between the right sides of (39) and (40) yields

$$L = h/(3^{1/2}\, m_n\, c). \tag{41}$$

In obtaining (41) we used that $\tau = L/c$.

By taking $m_n = 1.67 \times 10^{-27}$ Kg, we get

$$L = 0.76 \times 10^{-15}\, m = .76\, fm. \tag{42}$$

The above value can be compared with .81 fm, the value of the nucleon(proton) radius, as quoted in Halzen and Martin, ch. 8, [7].